\documentclass[reprint,superscriptaddress,showpacs]{revtex4}
\usepackage{epsf,amssymb,subfig,amsmath,color,times,natbib}
\usepackage[pdftex]{graphicx} 
\usepackage{amsmath}
\usepackage{mathtools}

\begin{document}
\title{Instability in electromagnetically driven flows\\
  Part I}

\author{Christophe Gissinger, Paola Rodriguez Imazio, Stephan Fauve}
  \affiliation{Laboratoire de Physique Statistique, Ecole Normale
  Superieure, CNRS, 24 rue Lhomond, 75005 Paris, France}

\newcommand{\bfnabla}{\boldsymbol{\nabla}} 
\newcommand{\bB}{\boldsymbol{B}}                   
\newcommand{\bj}{\boldsymbol{j}}                     
\newcommand{\bu}{\boldsymbol{u}}                  
\newcommand{\btimes}{\boldsymbol{\times}}   
\newcommand{\remark}[1]{{\color{red}\bf #1}} 

\begin{abstract}
The MHD flow driven by a travelling magnetic field (TMF) in an annular
channel is investigated numerically. For sufficiently large magnetic Reynolds number $Rm$, or if a large enough pressure gradient is externally applied, the system undergoes an instability in which the flow rate in the channel dramatically drops from synchronism with the wave to much smaller velocities. This transition takes the form of a saddle-node bifurcation for the time-averaged quantities. In this first paper, we characterize the bifurcation, and study the stability of the flow as a function of several parameters. We show that the bifurcation of the flow involves a bistability between Poiseuille-like and Hartman-like regimes, and relies on magnetic flux expulsion. Based on this observation, new predictions are made for the occurrence of this stalling instability.
\end{abstract}
\maketitle 

\section{Introduction}
\label{sec:intro}

The driving of an electrically conducting fluid by an electromagnetic
force due to a travelling magnetic field can yield significantly more complex behaviors than its solid
equivalent, the asynchronous motor. In particular, the many degrees of
freedom of the fluid and the presence of magnetohydrodynamic
boundary layers strongly affects the behavior of the system, and
several questions concerning the stability of such flows remain
unsolved.

 These questions become of primary interest in many industrial
 situations. For instance, Electromagnetic Linear Induction Pumps
 (EMPs) are largely used in secondary cooling systems of fast breeder
 reactors, mainly because of the ease of maintenance due to the absence
 of bearings, seals and moving parts. In these EMPs, the conducting
 fluid is generally driven in a cylindrical annular channel by means
 of a travelling magnetic field imposed by external coils. In such
 induction pumps, electrical currents in the liquid are induced by the
 variation of the magnetic flux of the wave, rather than injected through
 electrodes, as in conduction pumps \cite{Stelzer15}. In this perspective, several
 experimental studies of annular EMPs have been conducted
 \cite{Andreev78, Kliman79,Ota04}, in which the main
 characteristics of such pumps under large values of MHD interaction
 parameters have been described.\\

It is expected that as these pumps become large enough, a
magnetohydrodynamic instability arises, yielding significant decrease
in the developed flow rate. Gailitis and Lielausis \cite{Galaitis76} first
provided a theoretical analysis of this problem and proposed a
magnetohydrodynamic criterion for the occurrence of the instability in
such pumps. Based on the large aspect ratio $r_{inner}/r_{outer}$
generally used in experimental or industrial pumps, this theoretical
description of electromagnetically driven flows relies on the
assumption that the velocity field is invariant in the radial
direction.  According to this model, the instability occurs above some
critical magnetic Reynolds number $Rm_c$ and takes the form of an
inhomogeneity in the azimuthal direction. Several experimental studies
\cite{Kirillov80,Araseki04} have shown that when $Rm>Rm_c$, a
low frequency pulsation in the pressure and the flow rate is indeed
observed, strongly reducing the efficiency of the pump.
More recently, significant progress has been done on the understanding
of these electromagnetically driven flows. First, it has been shown
through numerical and experimental studies \cite{Araseki00} that even
at low $Rm$, the efficiency of such pumps is affected by an
amplification of the electromagnetic forcing, which takes the form of
a strong pulsation at double supply frequency (DSF). Second, it has
been confirmed that at large $Rm$, an azimuthal non-uniformity of the
applied magnetic field or of the sodium inlet velocity can create
vortices in the annular gap \cite{Araseki04}. In both cases
(large and low $Rm$), some solutions have been proposed to inhibit the
occurrence of these perturbations, but this generally leads to a strong
loss of efficiency of the pump. One purpose of the present article is
to provide a study of the physical mechanism by which the flow becomes
unstable in such systems in order to suggest new types of solutions
for suppressing the occurrences of the perturbations mentioned above.\\

In the present work, we numerically investigate a simple configuration
aiming to reproduce an electromagnetic pump. In the first part of the
article, we report calculations in the laminar regime and describe the instability arising when the magnetic Reynolds number is
large enough. After emphasizing the physical mechanism by which the instability occurs, we propose a new criterion for the instability threshold.  
In the second part of the paper, a new type of
instability is reported in 2D calculations performed at large Reynolds
numbers. We show how large scale vortices are created in the flow due
to induction effect, but only when the flow is turbulent enough. We
show that this instability is related to end-effects and can be 
understood in the framework of a simple model. Finally, we discuss these results in the perspective of
large scale experimental electromagnetic pumps and propose a
configuration inhibiting these type of destabilization.

\section{Theoretical description of an ideal cylindrical machine}
\label{sec:theory}

Before describing our numerical results, it may be useful to briefly
recall the equations obtained in the framework of the traditional
theory of induction-type cylindrical magnetohydrodynamics (MHD) machines.

\begin{figure*}
\centerline{\includegraphics[width= 10 cm]{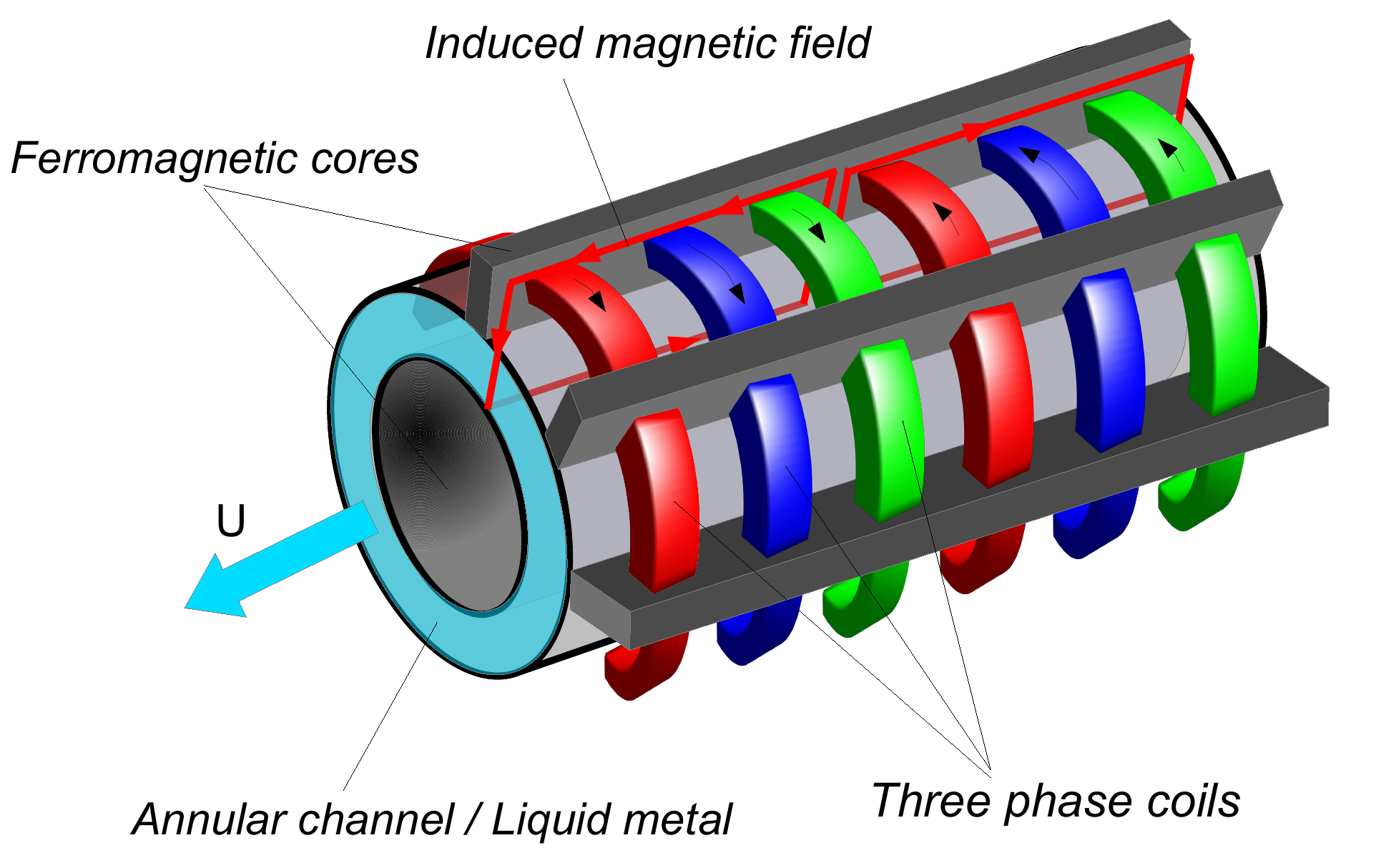} }
\caption{Schematic view of a typical electromagnetic pump}
\label{schema}
\end{figure*}

A schematic view of a typical electromagnetic pump is shown in
Fig. \ref{schema}. In general, the liquid metal flows along an annular
channel confined by two concentric cylinders. Some ferromagnetic cores
can be placed on the inner side of the channel, in order to reinforce
the radial component of the magnetic field created by a three-phase
winding of electrical currents imposed on the outer side of the
channel.\\

The governing equations are the magnetohydrodynamics equations,
i.e. the Navier-Stokes equation coupled to the induction equation

 \begin{equation}
 \rho\frac{\partial \bu}{\partial t} + \rho\left( \bu\cdot\bfnabla\right)  
\bu=-\bfnabla P+\rho\nu\bfnabla^{2}\bu +\bj\btimes\bB  \ ,
\label{NS}
\end{equation}
\begin{equation}
\frac{\partial\bB}{\partial t}=\bfnabla\btimes\left(\bu\btimes\bB\right)+\frac{1}{\mu_{0}\sigma}\bfnabla^2\bB  \ .
\label{ind}
 \end{equation}
where $\rho$ is the density, $\nu$ is the kinematic viscosity, $\sigma$,
is the electrical conductivity, $\mu_0$ is the magnetic permeability, $\nu$
is the fluid velocity, $\bB$ is the magnetic field, and $\bj=\mu_0^{-1}
\bfnabla\btimes\bB$ is the electrical current density. \\

In several references, electromagnetic pumps are modeled through a
modification of the above classical MHD equations solved in the liquid
domain, for instance by adding a source term in equation \ref{ind}
which represents the effect of the external coils. As this approach is
generally safe when modeling an external {\it homogeneous} and {\it
  stationary} magnetic field, the interpretation of such a
modification is more complicated when dealing with a travelling wave,
since it could be equivalent to add some unphysical source term to Maxwell's
equations. For instance, some authors incorrectly use a modified
Ohm's law in which a source term $J_0$ is added, and interpret $B$ in
equation \ref{ind} as an 'induced' magnetic field. 

Another approach is to decompose the total
magnetic field $B$  in the channel between
two parts $B=B_0+b$, where $B_0$ is the magnetic field that would be
created inside the annular gap {\it in absence of conducting media},
by any external sources (coils, permanent magnets, etc). $b$ can
therefore be regarded as the magnetic field generated by the {\it
  presence} of induced currents within the conducting fluid.\\

In the present article, we chose to use a third approach relying on
the modification of boundary conditions only, which ensures no
modification of Maxwell's equations whereas keeping a relatively simple
modelisation, without dealing with the external domain.

On the cylinders, we consider high permeability  boundary
conditions, for which the magnetic field is forced to be normal to
each boundary. This models the presence of ferromagnetic cores
generally found in electromagnetic pumps.  In addition, the presence
of azimuthal coils is modeled by an azimuthal electrical current
$J_s$ imposed on the outer cylinder, such that the boundary
conditions on $r_o$ become:

\begin{eqnarray}
\left({\bf B_2}-{\bf B_1}\right)\cdot{\bf e_r}&=&0\\
{\bf e_{r}}\times \left(\frac{{\bf B_2}}{\mu_2}- \frac{{\bf
    B_1}}{\mu_1}\right)&=&J_s{\bf e_\phi}
\end{eqnarray}

where the subscripts $1$ and $2$ respectively refers to the liquid metal (inside) and the external boundary (outside). In the limit of an external ferromagnetic boundary ($\mu_1/\mu_2\ll 1$),
in which the magnetic permeability $\mu_2$ of the boundary is much
larger than the permeability of the fluid $\mu_1$, this reduces to

\begin{eqnarray}
B_\phi(r_o)=0 \hspace{2mm}&,& \hspace{2mm}
B_z(r_o)=\mu_1J_s
\label{BC1}\\
B_\phi(r_i)=0  \hspace{2mm}&,& \hspace{2mm}
B_z(r_i)=0
\label{BC2}
\end{eqnarray}

The surface  current $J_s$ is
imposed :
\begin{equation}
J_s=J_0e^{i\left(kz-\omega t\right)}
\end{equation}
where $J_0$ is the complex amplitude of the applied current. Therefore,  $\omega$ and
$k=2\pi/\lambda$ are respectively the pulsation and the wavenumber of
the wave.\\

When considering an axisymmetric problem, one can  introduce the vector potential $A(r,z,t){\bf e_\phi}$ defined by ${\bf B}={\bf \nabla \times A} $, such that the induction equation becomes :

\begin{equation}
\frac{\partial A}{\partial t}+ V\frac{\partial A}{\partial z}=\frac{1}{\mu_{0}\sigma} \left( \frac{\partial^2A}{\partial^2 r} +  \frac{\partial^2A}{\partial^2 z}\right)
\label{potA}
\end{equation}

\noindent where the small-gap limit has been used.
Here, ${\bf V}=V{\bf e_z}$ is the velocity of the fluid, and in the
following theoretical description will be assumed to be only in the
$z$-direction, and completely homogeneous over the section of the
annular gap. Note that this homogeneous velocity of the fluid over the
entire cross-section constitutes one of the most drastic assumption of
the present derivation, and clearly contradicts the no-slip boundary
conditions used in our simulations. Due to the sinusoidal current applied on the outer boundary, we assume that the vector potential is of the form $A(r,z,t)=\tilde{A}(r,t)e^{ikz}$, leading to the equation:

\begin{equation}
\partial_t \tilde{A}+ ik\tilde{A}V=\eta(-k^2\tilde{A}+\partial_{rr}\tilde{A})
\label{A1}
\end{equation}
 
\noindent where $\eta=1/\mu_{0}\sigma$ is the magnetic diffusivity. This equation can be solved  by using boundary conditions (\ref{BC1})-(\ref{BC2}) and gives the radial structure of the magnetic field across the gap for a given velocity $V$.  Here, we will rather integrate this equation along the radial direction in the gap $d=r_o-r_i$. By defining the r-averaged vector potential $\Phi=\displaystyle \frac{1}{d} \int_{r_i}^{r_o} \tilde{A} \, \mathrm{d}r$, eq. (\ref{A1}) becomes:

\begin{equation}
\partial_t \Phi= -(ikV+\eta k^2)\Phi+ \frac{J_0}{d\sigma}e^{-i\omega t} 
\label{phi}
\end{equation}

\noindent where  the boundary conditions $\partial_r\tilde{A}|_{r_o}=J_0e^{-i\omega t}$ and $\partial_r\tilde{A}|_{r_i}=0$ have been used.\\

Similarly to $\Phi$, we can compute the large scale r-averaged induced current:
  
  \begin{equation}
  J(z,t)=\frac{1}{d}\left(\int_{r_i}^{r_o}j(r,z,t)\,\mathrm{d}r - J_s\right)=\left(\frac{k^2}{\mu}\Phi - \frac{2J_0}{d}e^{-i\omega t}\right)e^{ikz}
  \end{equation} 

The total  Lorentz force (averaged in the axial direction) acting on the fluid is then given by:

\begin{equation}
F_B=\frac{1}{2}\Re\{ {\bf J^*}\times {\bf B} \} = \Re\{i\frac{kJ_0}{d}\Phi e^{i\omega t} \}
\label{Fb}
\end{equation}

\noindent Here, both $J$ and $B$ in the above expression are large scale quantities averaged in the radial direction, implying that r-dependent perturbations are considered small compared to the mean values. Since the Lorentz force is quadratic in the magnetic field, it generally produces  a forcing on the flow with a component at a pulsation $2\omega$, which is
twice the one of the applied current. At small magnetic Reynolds number, this component can
be amplified and leads to the so-called Double Supply Frequency (DSF)
pulsation observed in some experimental pumps \cite{Araseki00}. Note that this component is filtered out in the present model, since only averaged quantities are considered.\\

In addition to this electromagnetic force, the flow undergoes a friction force due to the viscosity. If the Darcy-Weisbach equation for a pipe flow in the limit of purely hydrodynamical flows is used to evaluate the pressure loss due to friction, Navier-Stokes equations reduce to:

\begin{equation}
\rho\frac{\partial V}{\partial t}= K +F_B -\frac{\lambda\rho}{d} V^2 \text{sign}(V)
\label{DW}
\end{equation}

\noindent where $\rho$ is the density, $K=-\partial_zp$ is some imposed axial pressure gradient and  $\lambda$ is a friction parameter (supposed independent of $Re$ for simplicity). 

Finally, by taking a timescale $t_0=1/\omega$, a velocity scale $v_0=\omega/k=c$ and a scale of magnetic field $B_0=\mu J_0$, and writing eq. (\ref{phi}) and (\ref{DW}) in dimensionless form, we get the following system:

\begin{eqnarray}
R\dot{\Phi}&=& -(1+iRq)\Phi+e^{-iT} 
\label{dyn1}\\
\dot{q}&=& -\gamma q^2\text{sign}(q) - \frac{H^2\gamma}{R}\Re\{i\Phi e^{iT} \} + \kappa
\label{dyn2}
\end{eqnarray}

\noindent with $\kappa=\Delta P/\rho c^2$ and $\gamma=\lambda/k\delta$. The two dimensionless numbers  $R=\omega/\eta k^2$ and $H=\mu_0J_0^2/(\rho\lambda\omega\eta kd))$  can be regarded as respectively a magnetic Reynolds number and an Hartmann number based on the wave characteristics.  $q=v/c$ is the normalized velocity of the fluid and $T$ the dimensionless time. Equations (\ref{dyn1}) and (\ref{dyn2}) therefore provide a simple dynamical system describing the evolution of space averaged fields in the electromagnetic pump. \\

Let us now search for a pure travelling wave solution, for which $\dot{q}=0$ and $\Phi=\Phi_0e^{-iT}$. In this case, our system of equations simply reduces to:

\begin{equation}
\Phi_0=\frac{1}{1-iR(1-q)}
\end{equation}

\begin{equation}
\kappa=H^2\frac{1-q}{1+R^2(1-q)^2} -q^2\text{sign}(q)=F(q)
\label{PQ}
\end{equation}

\noindent The last equation is the so-called PQ-characteristic of the pump, in which the applied pressure gradient balances the Lorentz force and  the viscous force.  It is identical to the expression found by \cite{Galaitis76}.

\begin{figure}
\centerline{\includegraphics[width= 12 cm]{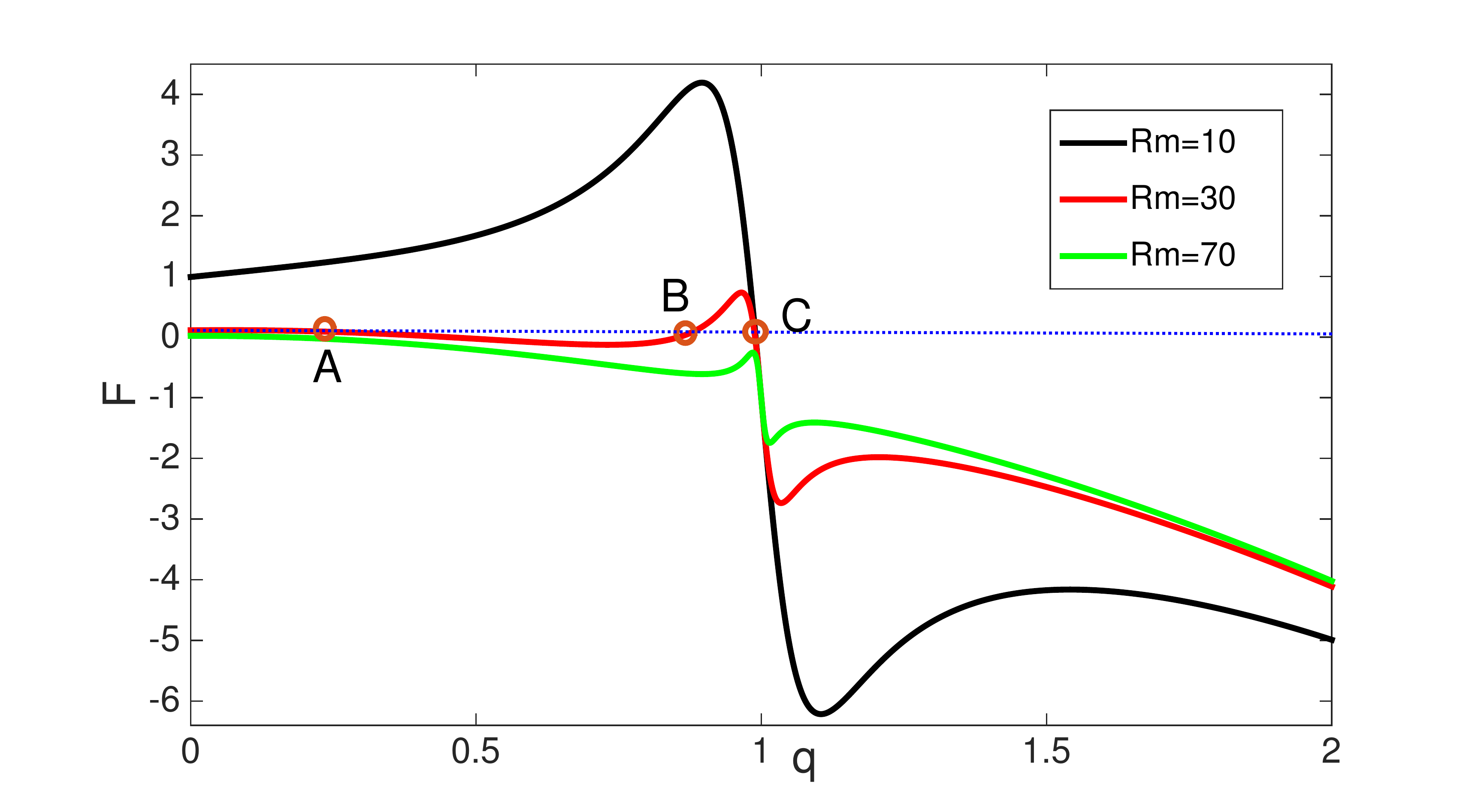} }
\caption{Theoretical characteristic curve F(q) for $Ha=10$, and
  $R=10$ (black curve), $R=30$ (red) and $R=70$ (green).}
\label{PQtheory}
\end{figure}

Due to the assumption of an homogeneous velocity, expression
(\ref{PQ}) is very similar to the equation predicting the stalling of
an asynchronous motor, and is sometimes referred as the PQ (for
Pressure P and Flow rate Q) characteristic of the pump. Figure
\ref{PQtheory} helps to understand the typical transitions prone to
appear in such systems. At large $H$ and moderate $R$, the viscous
dissipation is relatively small compared to the magnetic force, and
the system operates near $q=1$, where $F=0$ (for instance the black
curve in fig. \ref{PQtheory}). As $R$ is increased (red curve), one can
see that $F(q)$ intersects the line $F=0$ in three locations
corresponding to three different fixed points ($A$, $B$ and $C$). Two
of these solutions are stable equilibria (both points $A$ and $C$ lie
on the descending parts of the curve) whereas the third one is
unstable ($B$ is located on the ascending part). A further increase of
$R$ (green curve) leads to the coalescence of stable point $C$ and
unstable point $B$, and only the last stable solution $A$ remains. The
corresponding transition from equilibrium $C$ to solution $A$
therefore corresponds to a sudden drop of the total flow rate
developed by the system.\\

In the framework of equation \ref{PQ}, the coalescence of points
$B$ and $C$ corresponds to a saddle-node bifurcation between a stable
and an unstable {\it fixed points}. Note however that this
interpretation applies only to the time-averaged states of the
system.

This is an important difference with global numerical simulations or
experiments: in a real system, there are no stationary solutions since
the travelling magnetic field always induces some flow pulsation
through the Lorentz force. It is therefore not obvious that the
statistically steady states of a real pump with many degrees of
freedom will follow the same scenario than the simple saddle-node
bifurcation described here.\\

The above theory is based on many assumptions, such as an homogeneous
axial velocity, a simple travelling wave structure of the magnetic
field, or simplifications due to the small gap approximation.

In the next sections, we will first show to which extend this simple
description remains relevant to more complex systems such as numerical
simulations of laminar flows. In the last section, we will then
emphasize the physical mechanisms controlling the bifurcations of the
flow.

\section{Numerical model}
\label{sec:model}

In our numerical simulations, we consider the flow of an electrically
conducting fluid between two concentric cylinders. $r_i$ is the radius
of the inner cylinder, $r_o$ is the radius of the outer
cylinder, and $L$ is the length of the annular channel between the
cylinders. In all the simulations reported here, axial periodic boundary conditions are considered.

Among the different numerical studies aiming to model Annular Linear
EMPs, there are essentially two different ways to implement the effect
of the external windings.  A first approach consists in modifying the classical MHD equations
solved in the liquid domain, for instance by adding a source term in
equation \ref{ind} which represents the effect of the external
coils. As discussed in the previous section, this can lead to a
misleading interpretation on the magnetic field which is computed in
such models.  In the second approach, the external windings are explicitly modeled,
for instance through a finite element modelisation which is then
coupled to the MHD equations solved inside the fluid domain. In this
case, Maxwell's equations are not modified, and the modelisation can
be made very close to experimental configurations. This however
generally requires the introduction of several parameters (describing
the external windings) and the use of a complex numerical code based
on the  matching between the bulk flow and the external domain.\\

In the present paper, only boundary conditions are modified in the spirit of the above theoretical description, which ensures no modification of Maxwell's equations whereas keeping a relatively simple
modelisation, without dealing with the external domain. This approach is therefore very close to the theoretical model developed in the previous section.

On the cylinders, large magnetic permeability boundary conditions identical to eqs.(\ref{BC1}-\ref{BC2}) are applied. In our numerical simulations, the surface current at the outer boundary is imposed :

\begin{equation}
J=J_0\sin\left(kz-\omega t\right)
\end{equation}
where $J_0$ is the amplitude of the applied current,  $\omega$ and
$k=2\pi/\lambda$ are respectively the pulsation and the wavenumber of
the  wave.

Note that due to the solenoidality of the magnetic field, these
boundary conditions lead to the generation of a strong radial
magnetic field as well. Since both $B_z$ and $B_\phi$ vanish at
$r=r_i$, this approach models the generation of a strong radial
field $B_r$ propagating through the channel in the absence of
velocity.\\

In the code, the equations are made dimensionless using a  length scale
$l_0=\sqrt{r_i(r_o-r_i)}$ and a  velocity scale $U_0=c$, where
$c=\omega/k$ is the speed of the  wave. The
pressure scale  is $P_0=\rho c^2$ and the scale of the magnetic field is
$B_0=\sqrt{\mu\rho}c$.

The problem is then governed by the geometrical parameters
$\Gamma=L/(r_o-r_i)$ and $\beta=r_i/r_o$, the magnetic Reynolds number
$Rm=\mu_0\sigma cl_o$ and the magnetic Prandtl number
$Pm=\nu/\eta$, where $\eta=1/\mu_{0}\sigma$. Note that here, $Rm$ is based on the wave speed $c$ rather than the fluid velocity. Using this definition, $Rm$ is a given control parameter and not an output of the numerical simulations. The magnitude of the imposed current is controlled by
the Hartmann number, defined as
$Ha=\mu_0J_0l_0/\sqrt{\mu_0\rho\nu\eta}$. Alternatively, one may define the kinetic Reynolds number $Re=Rm/Pm$ instead of $Pm$. Note that both $Rm$ and $Re$ are defined here using the speed of the wave and not the
one of the fluid.  In our simulations, $Pm$ is varied between $6.10^{-3}$ and $10$, whereas real liquid metals have $Pm\leq 10^{-5}$.

These equations are integrated with the massively parallel HERACLES code
\cite{Gonzales06}. Originally developed for radiative astrophysical
and ideal-MHD flows, it was modified to include viscous and magnetic
diffusion. Note that HERACLES is a compressible code, whereas
laboratory experiments generally involve almost incompressible liquid
metals. In fact, incompressibility corresponds to an idealization in
the limit of infinitely small Mach number ($Ma$). In the simulations
reported here, we used an isothermal equation of state with a small
sound speed (in practice $Ma=0.03$), following the approach of
\cite{Liu06,Liu08,Gissinger11}. Typical resolutions
used in the simulations reported in this article range from
$(N_r,N_Z)=[128,512]$ for laminar flows to $(N_r,N_Z)=[256,1024]$ for
the highest Reynolds numbers, computed typically on $256$ processors. 
All the simulations reported here has been evolved  up to $t=1000$. For the velocity field, no-slip conditions are used at the radial boundaries. Depending on the simulations, we can either impose an inlet velocity $U_{in}$ at $z=0$,
or an applied pressure gradient $P_{in}$ over the whole pump. 

\section{Numerical simulation: stalling instability at large Rm}
\label{sec:num}

\begin{figure*}
\centerline{\includegraphics[width= 16 cm]{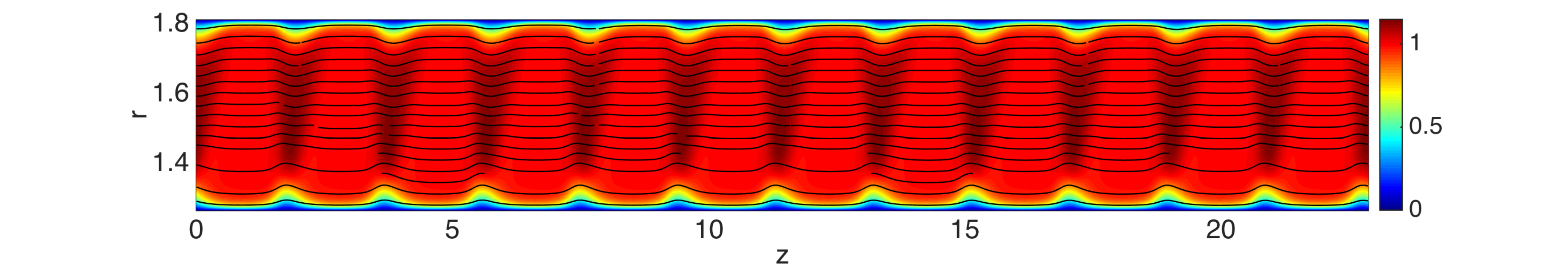} }
\centerline{\includegraphics[width= 16 cm]{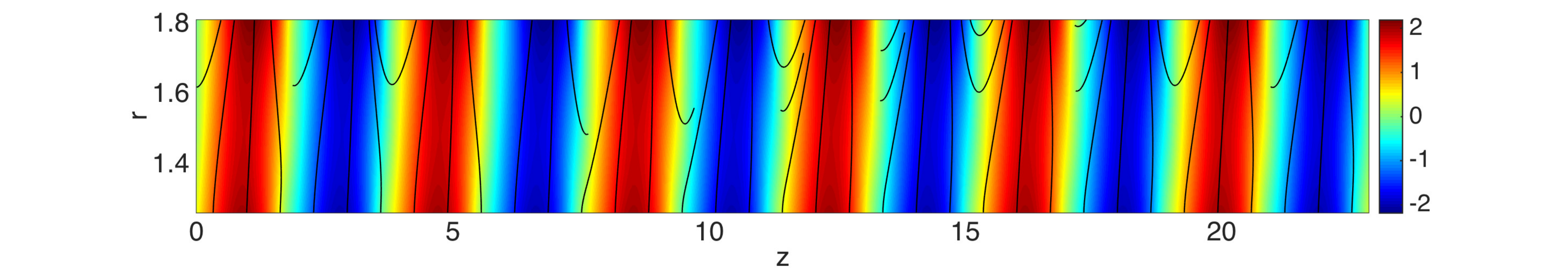} }
\caption{Structure of the axial velocity (top) and the radial magnetic
  field (bottom) for $Rm=30$, $Ha=100$ and $Re=100$. Both velocity
  streamlines and magnetic field lines are indicated. Note that the fluid is pumped nearly in synchronism with the traveling magnetic field, similarly to the solution A in Fig. \ref{PQtheory}.}
\label{lamstruc}
\end{figure*}

In this section, we describe the results obtained with direct
numerical simulations. All the results reported here are performed in
axisymmetric configurations. 
Figure \ref{lamstruc} shows a typical run in the laminar case, for
$Rm=30$, $Re=100$ and $Ha=100$. The figure shows colorplot of the
axial velocity field $U_z$(top) and the radial magnetic field $B_r$
(bottom).

As expected, the external currents, applied only at $r=r_o$, generate
a strong radial magnetic field $B_r$. The extension of $B_r$ through
the whole channel is clearly reinforced by the ferromagnetic boundary
conditions on the inner cylinder at $r=r_i$, which ensures
$B_\phi=B_z=0$. Because of the solenoidal nature of the magnetic field
and its wavy structure in $z$, it also shows significant components in
the axial direction, despite the use of ferromagnetic
boundaries. Therefore, some of the magnetic field lines starting from
$r_o$ connects to $r_i$ whereas others reconnects directly to
$r_0$. Note that this complex structure of the field is not considered
in the basic theoretical model described in the previous section. As
this magnetic structure propagates in time in the $z$-direction, it
produces a Lorentz force which pumps the fluid mainly in the axial
direction, although the velocity streamlines exhibit some waviness
due to the presence of local forces in the radial direction.
Except for this waviness, the flow exhibits a relatively homogeneous structure, and is pumped in nearly synchronism with the traveling magnetic field. Note that this state is similar to the stable solution A (Fig. 2) of the simplified model presented in the previous section .\\

\begin{figure*}
\centerline{\includegraphics[width= 8 cm]{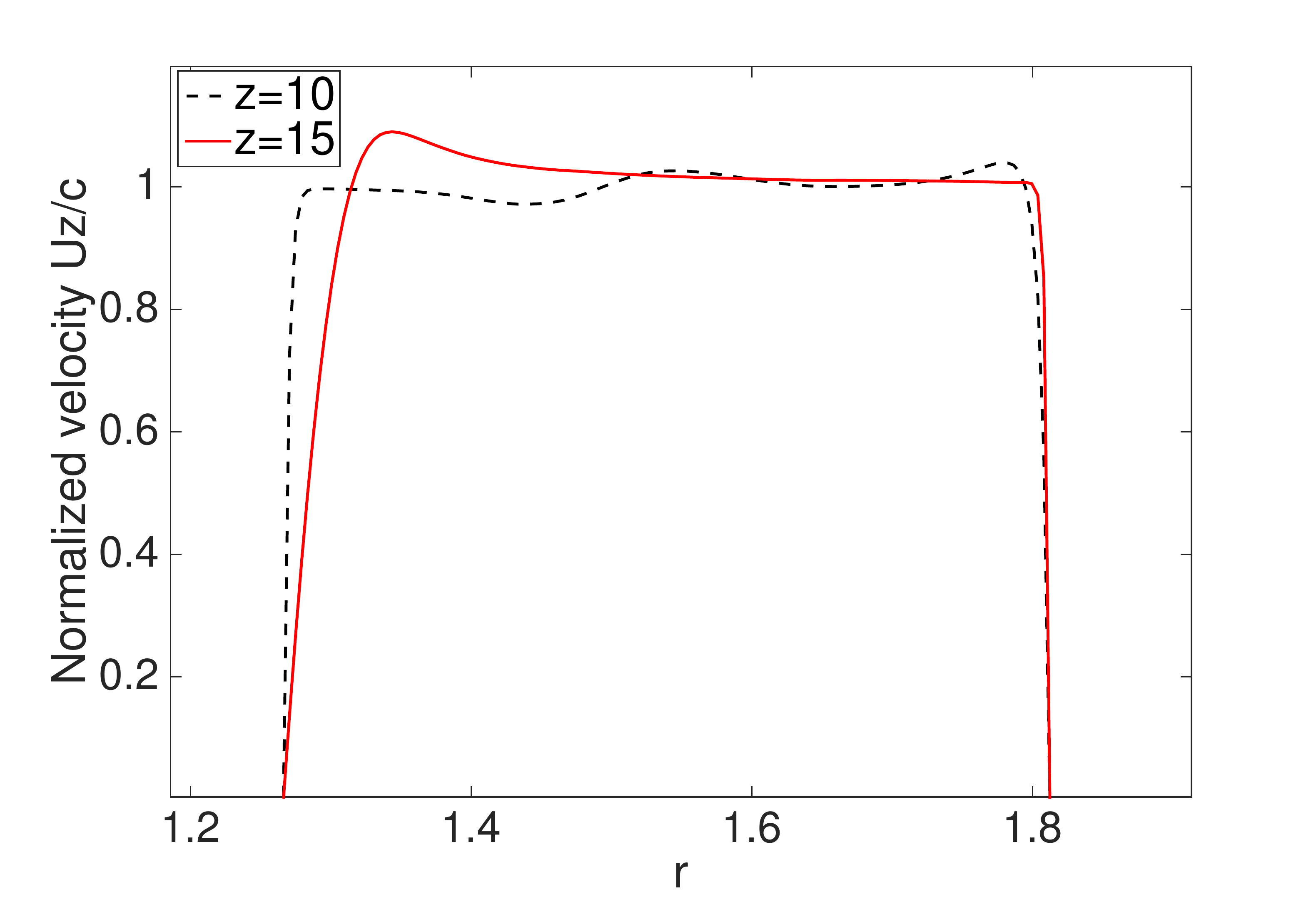} 
  \includegraphics[width= 8 cm]{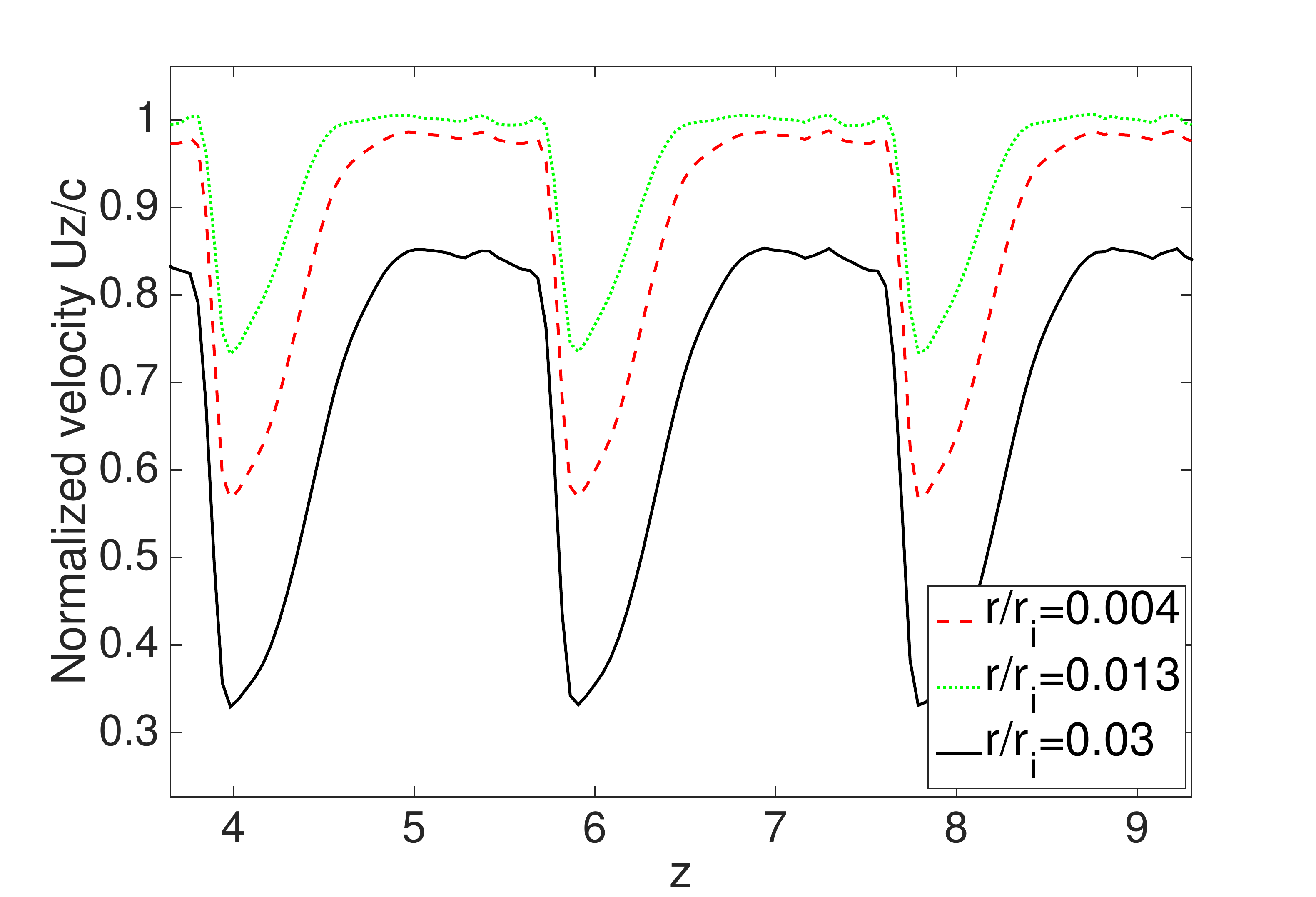} }
\caption{Profiles of the axial velocity field as a function of $r$
  (left) and as a function of $z$ for $Rm=30$, $Ha=100$ and
  $Re=100$. Note the difference with the classical Hartmann boundary
  layers.}
\label{profiles}
\end{figure*}

Figure \ref{profiles}-left shows instantaneous flow profiles as a
function of $r$ for two different values of $z$ (note that the same
would be obtained for different times at a given $z$ due to the
propagative nature of the magnetic field). At $z=1.5$ for instance, the profile
is very close to an Hartmann solution, in which the velocity $U_z$
(normalized with the wave speed $c$) is nearly equal to $1$ in the whole
cross-section. In other words, the flow is pumped in synchronism with
the travelling magnetic field. Close to the boundaries however, the flow is
forced to vanish due to the no-slip boundary condition, which induces
the presence of thin magnetic boundary layers at $r_i$ and $r_o$. It
is important to mention that these layers {\it are not} Hartmann
boundary layers as generally defined, since the magnetic field varies
both in space and time. For instance, as we move to larger $z$, the
shear close to the inner cylinder can be strongly reduced: the profile
computed at a different location $z=0.4$ (red curve) exhibits a
boundary layer near $r_i$ roughly $6$ times larger than for
$z=1.5$. It is also interesting to note that unlike the imposed
magnetic field which creates it, this variation of the boundary layer
thickness is not sinusoidal in $z$. Fig. \ref{profiles}-right shows
$U_z$ as a function of $z$ for different distances from the inner
cylinder. It illustrates the strong variation of the thickness of
these boundary layers, which consist of large regions where the
boundary layer is very thin, alternating with short events of
thickened layers (where $B_r$ changes sign). It is known that Hartmann
layers are linearly stable against axisymmetric disturbances but can
be destabilized in $3D$ calculations from transient growth of the
two-dimensional perturbations \cite{Krasnov04}. Here, one may expect
that the presence of velocity gradient along $z$ due to the variation
of the boundary layer thickness makes the flow more prone to boundary
layer instability. In particular, the thickness of these layers do not follow the Hartmann scaling $\delta\sim 1/Ha$ and strongly depends on $Rm$ as well. Although it is beyond the scope of this paper, it
would be interesting to compare the stability of the magnetic boundary
layers described here with classical results on Hartmann layers.\\

\begin{figure}
\centerline{\includegraphics[width= 10 cm]{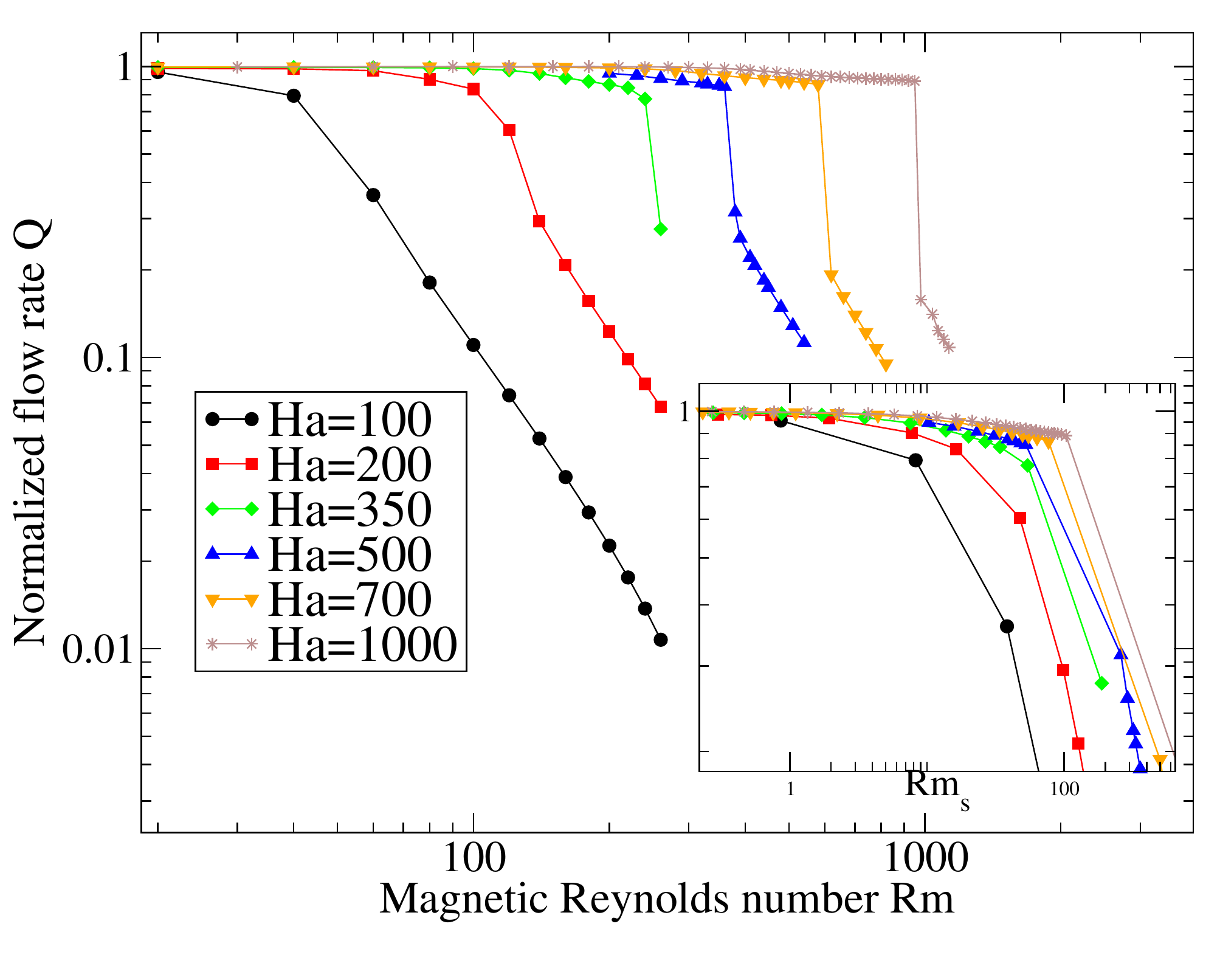} }              
\caption{Bifurcation of the axial flow rate as a function of $Rm$, for
  different values of $Ha$. Note the sharp transition observed at
  large $Ha,Rm$. Inset: same, but plotted as a function of the slip magnetic Reynolds number (see text))}
\label{bifRm}
\end{figure}

In fact, the fluid velocity strongly depends on the value of the
magnetic Reynolds number $Rm$ and the Hartmann number $Ha$. We have
performed numerical simulations with different values of $Rm$ and
$Ha$, at a fixed value of the kinetic Reynolds number $Re=100$,
in order to stay in the laminar regime. At this $Re$, the flow always
stays axisymmetric. For each simulation we computed the normalized
flow rate through the annular cross-section $S$ of the channel
$Q=\frac{1}{S}\int_{S}U_zdS$. Figure \ref{bifRm} shows the evolution of
the time averaged flow rate as a function of $Rm$, for different
values of $Ha$. For the smallest value of the Hartmann number
($Ha=100$, black circle curve), increasing $Rm$ yields a continuous decrease
of the fluid velocity, from nearly synchronism ($Q=0.95$) to very small
values ($Q\sim 0.02$ at $Rm=200$). Indeed, as $Rm$ is increased at
fixed $Re$, the ratio between the Lorentz force and the viscous
dissipation decreases, leading to a weaker driving of the flow. The
same occurs at low $Rm$ when $Ha$ is decreased. Note that these smooth
transitions show no hysteresis, since only one solution seems to be
involved. \\

For larger values of the Hartmann number (for instance $Ha=500$, blue
curve), the situation clearly differs: as $Rm$ is increased from $20$
to $350$, the flow is first only weakly modified, from $Q=0.95$ to
$Q=0.86$, the velocity of the fluid keeping values relatively close to
the wave speed. In this regime, the fluid velocity and
the magnetic field have the structure shown in Fig. \ref{lamstruc},
i.e. of Hartmann-type. However, around $Rm\sim 300$, the flow
undergoes a sharp transition from nearly synchronism with the wave to
much smaller velocities ($Q=0.25$). During this transition, the total
Lorentz force on the fluid also strongly decreases, and the large
difference of velocity between the fluid and the wave produces a
strong ohmic dissipation. This transition is therefore similar to the saddle-node bifurcation predicted in the model presented in section 2.

This curve can be compared to the calculation from \cite{Galaitis76}, which predicts that unstable flows in such pumps may arise when the magnetic Reynolds number based on the slip $Rm_s=Rm(1-q)$ becomes sufficiently large. In the inset of Fig.\ref{bifRm},  we have plotted the bifurcation diagram as a function of $Rm_s$. Although this slip Reynolds number  seems to slightly rescale our curves, it is clear that it cannot be used as the only rescaling parameter, since the stalling of the pump still depends on the value of the Hartmann number. Note also that the critical values of $Rm$ and $Rm_s$ are relatively large in this laminar regime, compared to the prediction $Rm_s>1$. We will see in the second part of the paper that these critical numbers strongly decrease with the fluid Reynolds number $Re$, bringing the instability pocket in the accessible parameter range of experimental pumps.

\begin{figure}
\centerline{\includegraphics[height= 5 cm]{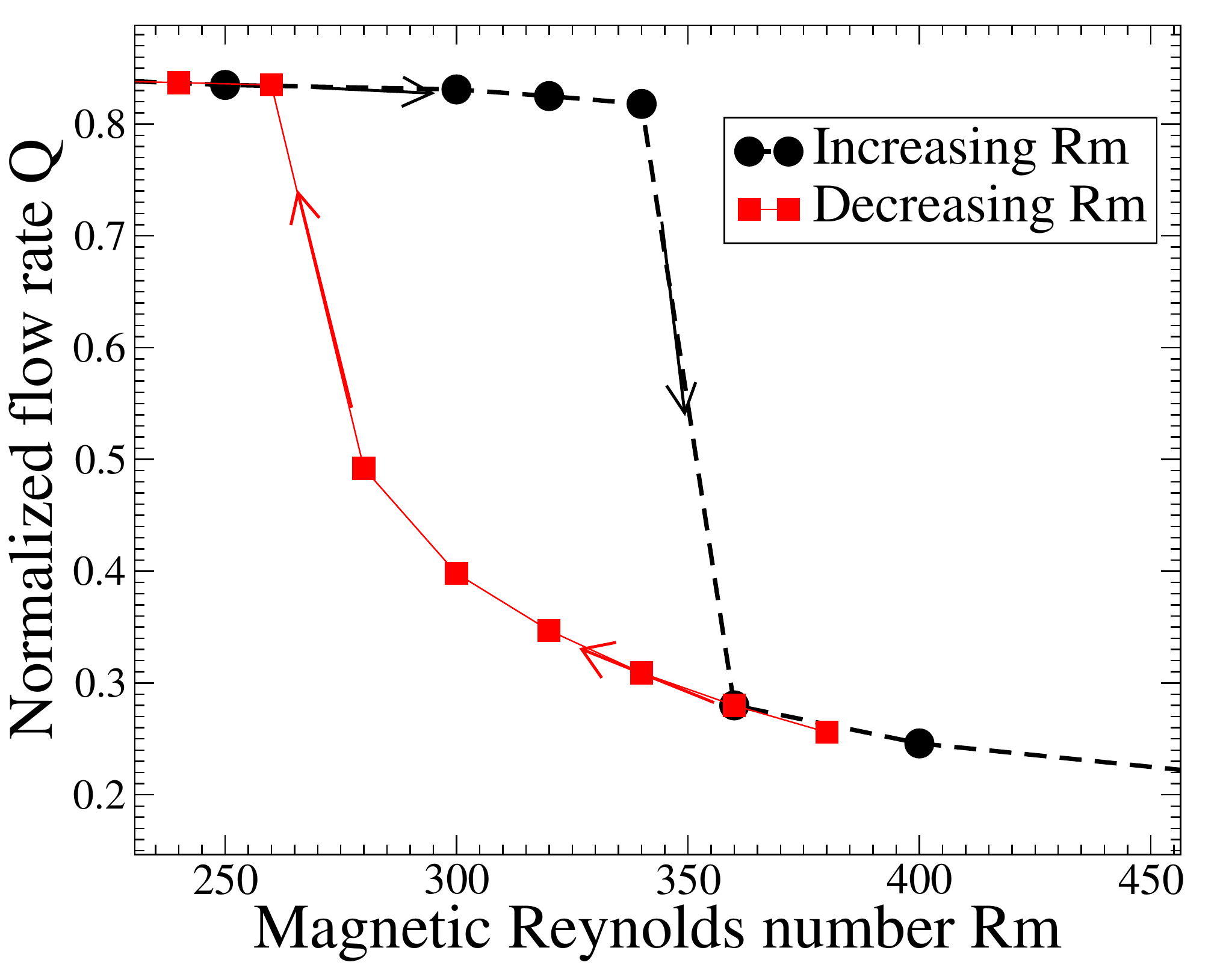} 
  \includegraphics[height= 5.2 cm]{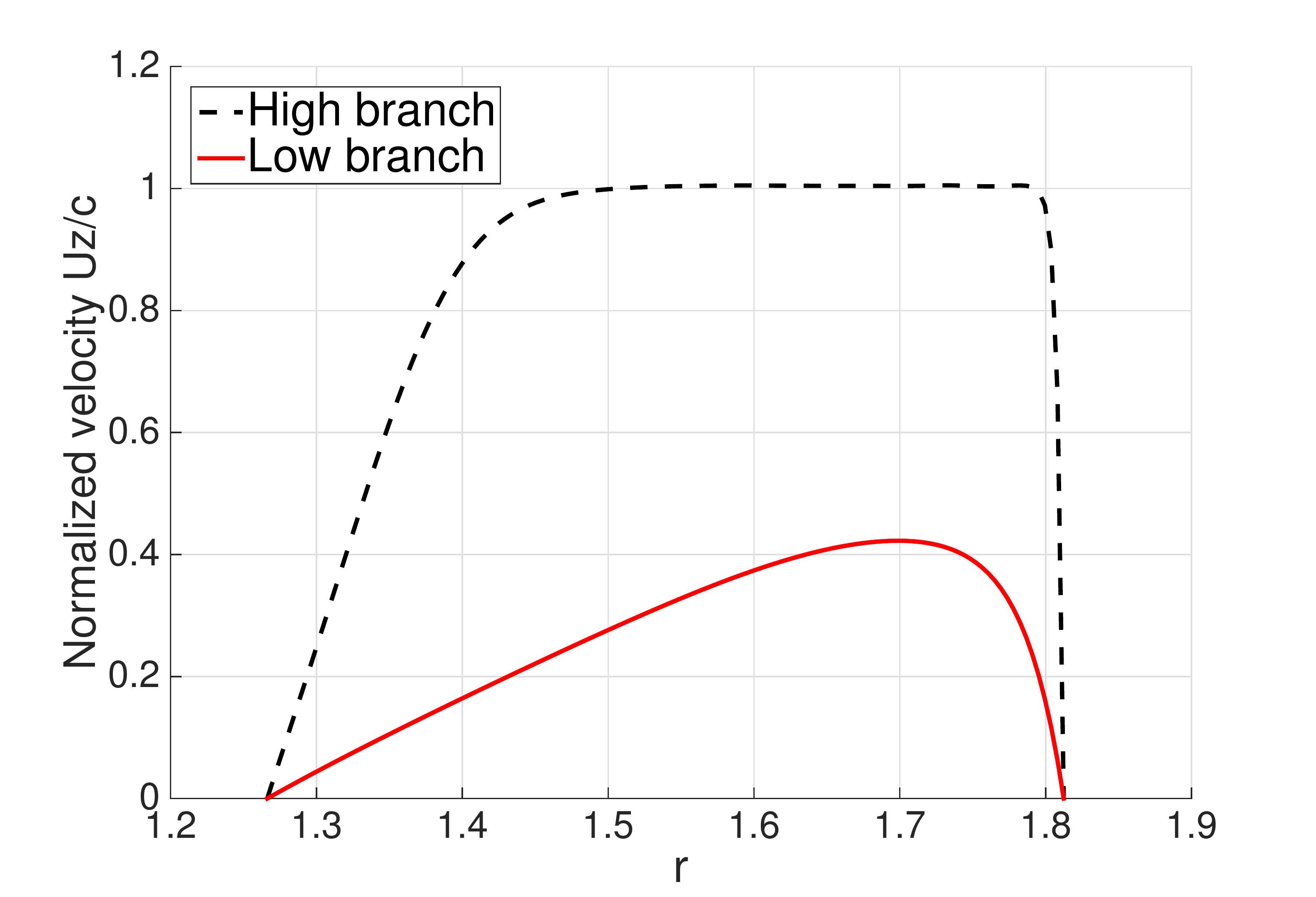} }
\caption{Left: Bifurcation diagram of the axial flow rate as a
  function of $Rm$, for $Ha=500$. Bistability between two solutions
  is observed, related to strong hysteresis. Right: radial
  profiles of the two solutions for $Rm=350$. }
\label{hysteresis}
\end{figure}

This transition is shown in more details in Fig. \ref{hysteresis},
which focuses on the transition observed at $Ha=500$. Black circles
correspond to simulations restarted from the calculation performed at
a smaller $Rm$, whereas red squares are obtained by continuing
simulations performed at larger $Rm$. Depending on initial conditions, the large speed and low speed solutions are
bistable on a relatively large range of the magnetic Reynolds number
$Rm$ (between $Rm=300$ and $Rm=390$ at $Ha=500$). The transition from
synchronous to stalled flows is therefore strongly subcritical, with a
strong hysteresis. Note that the bistability involves two solutions
with very different spatial structures. In particular, the synchronous
solution is associated to strong shear close to the boundaries,
but these Hartmann-like boundary layers almost disappear in the
stalled regime (see Fig.~\ref{hysteresis}-right), which is closer to a
Poiseuille flow.

\begin{figure*}
\centerline{\includegraphics[width= 8 cm]{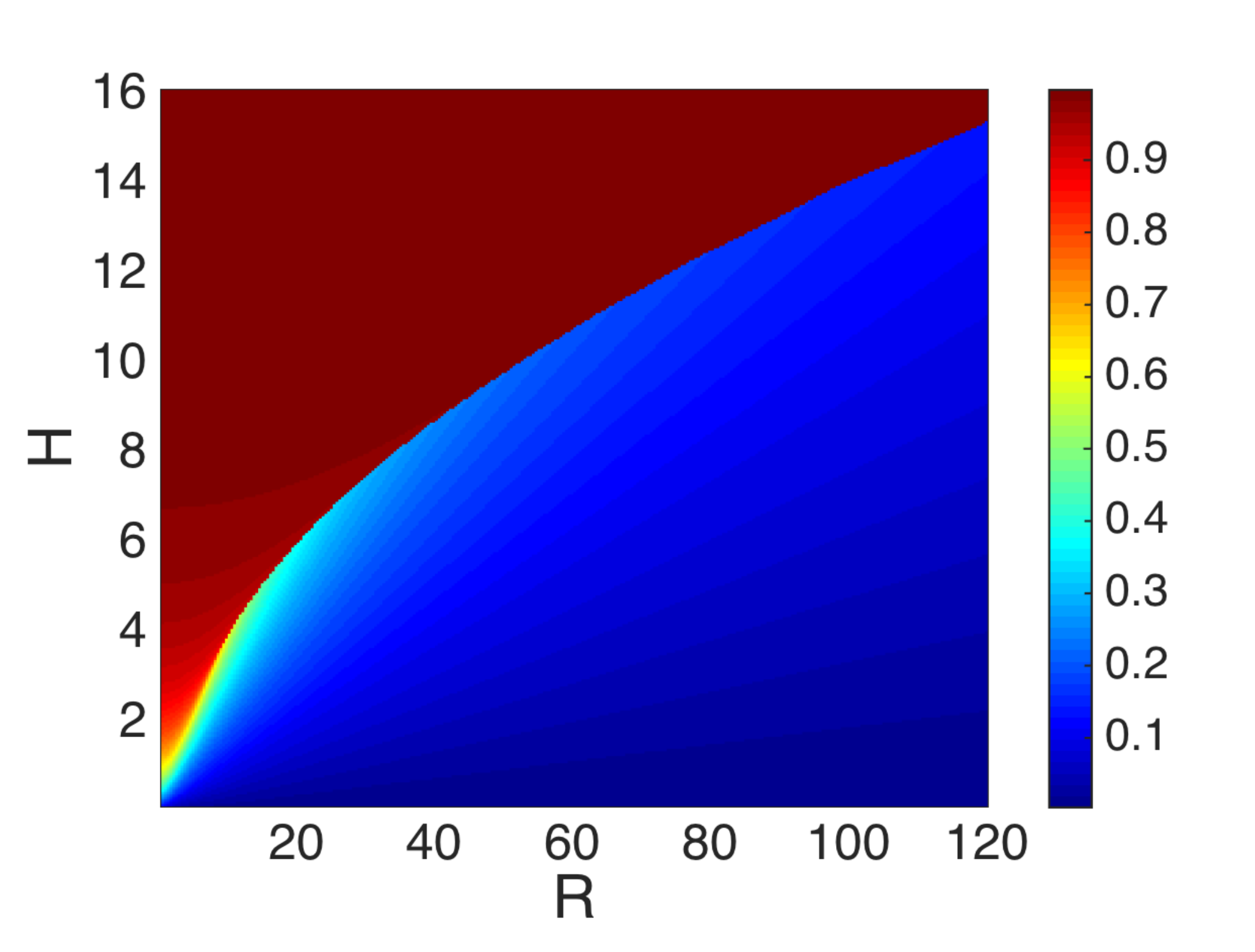} 
  \includegraphics[width= 8 cm]{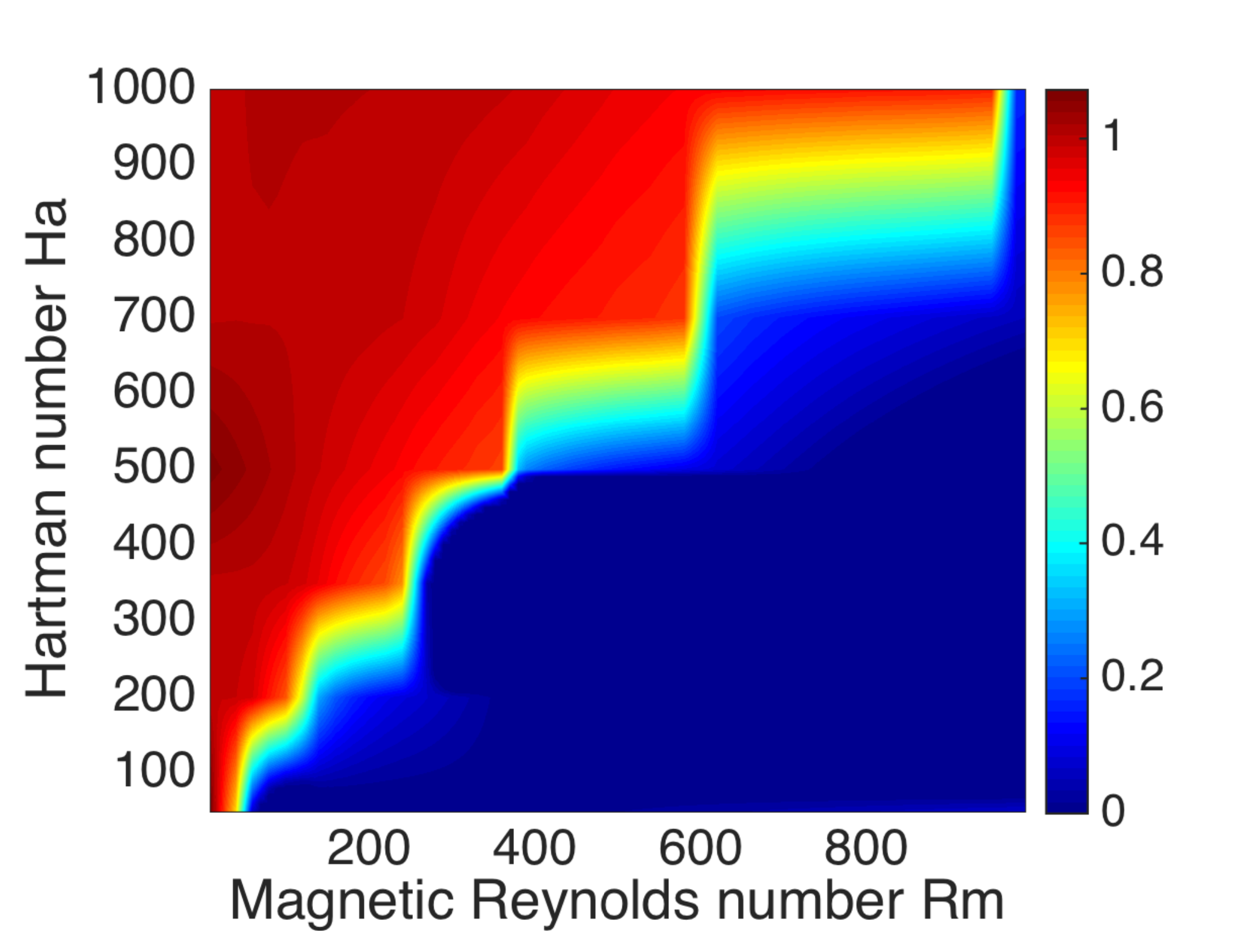} }
\caption{Amplitude of the axial flow rate $Q$ as a function of $Rm$
  and $Ha$. Left: theoretical prediction from eq. \ref{PQ}. Right:
  Global numerical simulations.}
\label{PQcompare}
\end{figure*}

Figure \ref{PQcompare} summarizes the stability of the flow by showing
a colorplot of $Q$ in the parameter space $(Ha-Rm)$. It compares the
velocity obtained in our numerical simulations for $Re=100$,
interpolated from results of Figure \ref{bifRm}, and predictions from
the expression given in equation \ref{PQ} (where only the solution at
larger $q$ is computed). All the simulations reported in this figure
have been obtained  at fixed $Ha$ and {\it
  increasing} $Rm$. It is interesting to note that in both theory and
simulations, the marginal stability curve follows $Ha_c\propto
\sqrt{Rm}$.  Because of the bistability described in
Fig.~\ref{hysteresis} and the subcritical nature of the transition,
note that a different picture is obtained by {\it decreasing} $Rm$ at
fixed $Ha$ (and taking the smallest solution of eq. \ref{PQ}). In this
case, one observes a critical Hartmann number to follow $Ha_c\propto
Rm$ instead.

Despite the strong assumptions made during the derivation of the
analytical model of the previous section, one can see that the
behavior of our electromagnetically driven flow in the laminar regime
stays very close to theoretical predictions. In particular, note that
although strongly irrelevant close to boundaries, this model based on
the assumption of an homogeneous velocity in $r$ correctly predicts
the shape of the marginal stability curve. This can be understood as a
consequence of the Hartmann-type profile observed at the onset of the
instability: first, the bulk flow is nearly constant in $r$, in
agreement with the theory. Second, since the magnetic boundary layers
remain stable (at least for the range of parameters explored here),
they do no play a significant role in the transitions, and the
dynamics is mainly dominated by the bulk flow only.

However, a closer inspection of the behavior of our simulations shows
that not only the magnitude but also the radial structure of the
velocity is strongly modified beyond the transition. This modification
of the radial profile suggests that the physical mechanism by which
the flow becomes unstable may not be captured by the simple
theoretical model of section \ref{sec:theory}. In particular, there is
no simple explanation for the stability line obtained in
Fig. \ref{PQcompare}. In the next section, we therefore investigate a
different type of simulations aiming to clarify the physics of the
stalling instability observed in electromagnetic pumps.

\section{Magnetic flux expulsion}
\label{sec:expulsion}

In a real pumps, there is always some pressure loss due to the friction
along the pipe, which can in general be related to the averaged
velocity of the fluid. In addition to this viscous dissipation, the
connection of the pump with the external hydraulic machinery can
sometimes induce an additional load acting on the system.

In this section we report another set of simulations, in which both
$Rm$ and $Ha$ are kept fixed, but an adverse pressure gradient,
modeling an external load, is applied to the pump. Although such a
load can have a complex dependence with the velocity, we choose here
to use a constant pressure gradient, independent of the flow. These
simulations are therefore closer to a real electromagnetic pump, in
which the physical properties of the pump are fixed but a variable
load is applied on the system. Note that the pressure gradient $P_{ext}$ is now the control parameter for the instability. From an industrial point of view, $Rm$ is still an important control parameter, since its critical value for the stalling instability corresponds to a critical size for large scale induction pump beyond which the flow pumping becomes inefficient.

\begin{figure}
\centerline{\includegraphics[width= 10 cm]{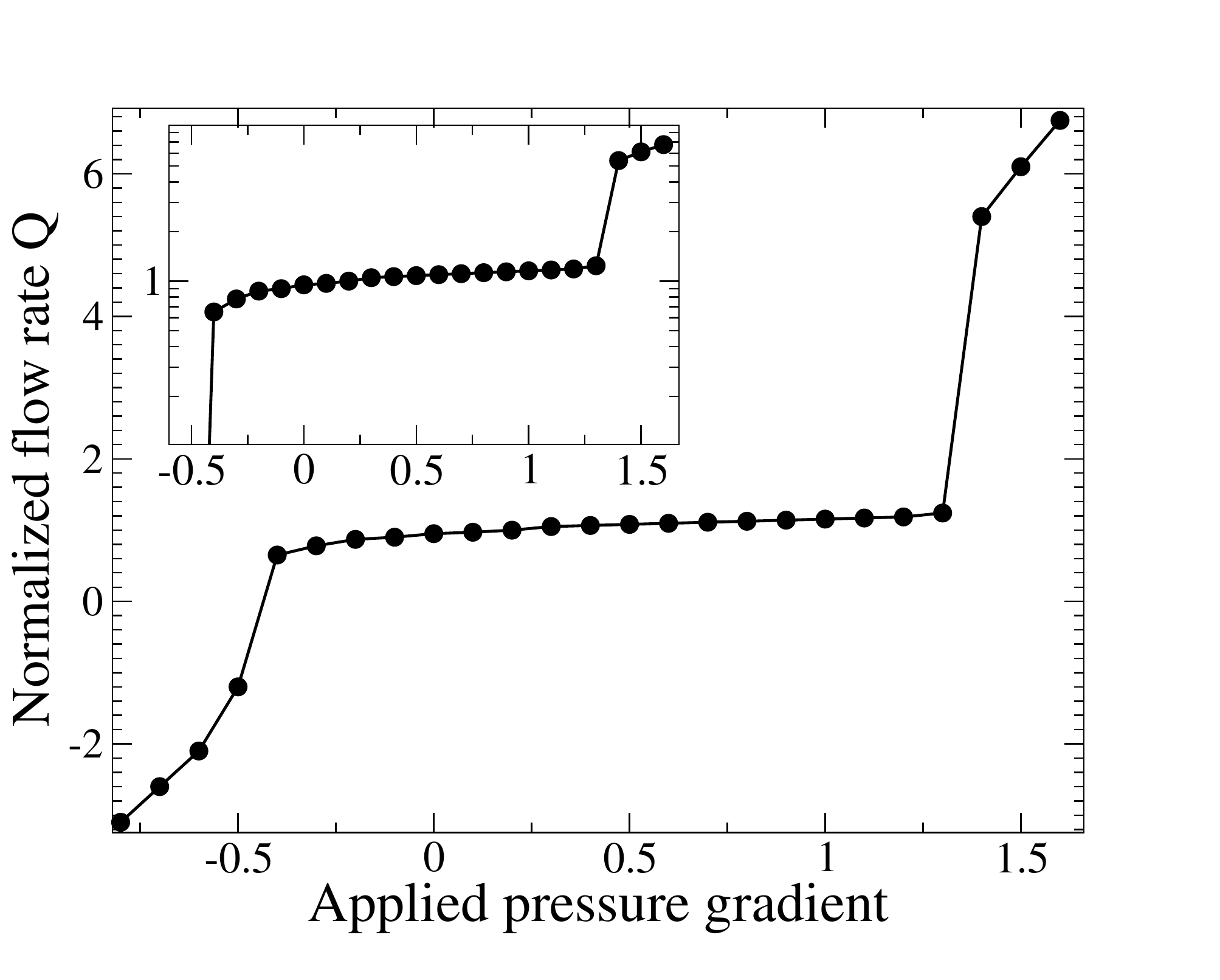} }
\caption{Stalling of the flow when submitted to an external load
  $P_{ext}$, for $Rm=120$ and $Ha=300$. For sufficiently large pressure gradient (in absolute
  value), the flow can suddenly either accelerate or change
  direction. Inset: same in log scale. }
\label{decrochA}
\end{figure}

The results are shown in Figure \ref{decrochA}, where the
time-averaged normalized flow rate of the fluid is plotted as a
function of the amplitude $P_{ext}$ of the load, for $Rm=120$ and
$Ha=300$.
In this figure, the black dots have been obtained by first performing one
simulation at $P_{ext}=0$, and then following this stable branch
towards positive and negative $P_{ext}$. In practice, this means that
for each value of $P_{ext}$, we used the previous simulation
(performed at smaller $|P_{ext}|$) as an initial condition.

Over a large range of $P_{ext}$, the system is relatively independent
of the value of the load, and the fluid moves in synchronism with the
wave, at a flow speed always larger than $0.8$. For a critical value
around $P_{ext}=-0.5$, the flow suddenly bifurcates to a very
different regime in which the fluid velocity is pumped backward to the
direction of the electromagnetic driving. In addition, this negative
velocity strongly depends on the value of the pressure gradient.

Interestingly, an identical transition, but towards large speed flows,
occurs if an accelerating pressure gradient is applied to the
pump. Note also, that once this bifurcation takes place, the evolution
of $Q$ also strongly depends on the pressure gradient. \\

\begin{figure}
\centerline{\includegraphics[width= 10 cm]{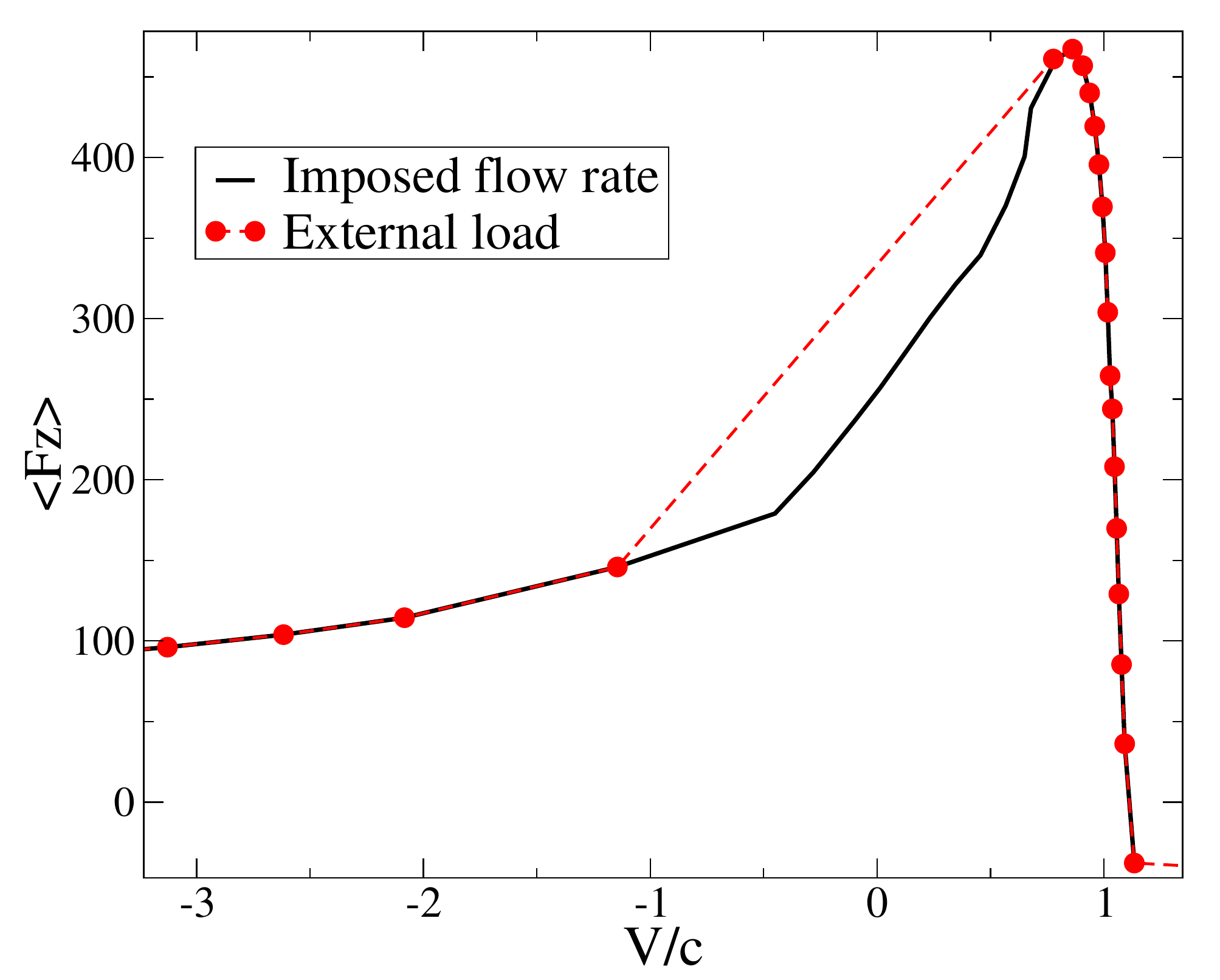} }
\caption{Evolution of the Lorentz force with the flow rate through the
  annular section. Black line: simulations using an imposed inlet
  velocity $U_{in}$ (see text). Red dots: simulations performed with
  an imposed pressure gradient $P_{ext}$, for the same parameters
  $Rm=120$ and $Ha=300$. The dashed line indicates the jump corresponding to the bifurcation. This PQ characteristic curve is  similar to the one predicted from the model in section 2 (see Fig.\ref{PQtheory}).}
\label{decrochB}
\end{figure}

The next figure helps to understand how this transition is related to
the stalling instability described in previous sections. In
Fig. \ref{decrochB}, the black curve has been obtained by performing
simulations in which the flow rate inside the channel is fully
determined by some inlet flow rate $Q_{in}$ which is imposed at
$z=0$. Since the flow is nearly incompressible, this flow rate is
conserved in the whole domain.

By varying $U_{in}$ and reporting the corresponding values of the
space and time averaged Lorentz force, we therefore obtain the
PQ-characteristic curve of the modeled pump. This curve compares very
well to the theoretical prediction shown in section
\ref{sec:theory}. As expected, the Lorentz force is nearly zero for
$Q=1$, exhibits a very clear maximum before synchronism, and rapidly
decreases as the velocity goes away from $Q=1$. Contrary to the
theoretical description, the averaged Lorentz force does not vanishes
as the velocity goes to strongly negative values. This can be
understood as a direct consequence of the no-slip boundary conditions and
the associated Hartmann-type boundary layers: close to the boundaries,
there is always some region for which $Q$ is close to zero, which
corresponds to a strong accelerating Lorentz force. \\

The red dots of Fig. \ref{decrochB} correspond to simulations shown in
Fig. \ref{decrochA}, i.e. with an applied pressure gradient $P_{ext}$
rather than an imposed velocity. When $P_{ext}=0$, the only forces
applied to the systems are the Lorentz force and a weak dissipation:
the pump therefore operates just below synchronism with the wave, the
difference being due to the presence of viscous dissipation. As
$P_{ext}$ is decreased from $0$ to $-0.5$, this additional external
load brakes the flow and pushes the system to lower values of $Q$,
which moves on the descending part of the black curve. Note that all
the points located on this descending branch $(\partial F_B/\partial
q<0)$ correspond to stable points, and the large (negative) slope of
this branch explains why strong variations of the external load only
weakly modify the value of the flow rate. Note however that the
electromagnetic force is strongly enhanced during this phase.

For $P_{ext}=-0.5$ however, when the system reaches to maximum value
of $F_B$, this {\it 'synchronous solution'} looses its stability and the
system bifurcates to another stable fixed point corresponding to
strongly negative values of the velocity, where the pressure gradient
and the viscous dissipation balance the Lorentz force.This new state
corresponds to a Poiseuille-type flow mainly imposed by the negative
pressure gradient, and only weakly affected by the Lorentz force
(compared to Fig. \ref{lamstruc}). The waviness of the fluid (and
therefore the effect of the imposed magnetic field) is still present
but its amplitude is negligible compared to the mean value.\\

From a physical point of view, this transition can be regarded as an
instability triggered by {\it magnetic flux expulsion}. When operating
on the descending part of its characteristic shown in
Fig. \ref{decrochB}, the pump lies on a stable fixed point, in which
the strong Lorentz force balance the force due to the adverse pressure
gradient: the mean flow is of Hartmann type. Since a decrease of the flow speed is associated to an increase of the accelerating Lorentz force, this branch is always stabilized by the magnetic tension.

\begin{figure*}
\centerline{\includegraphics[width=16 cm]{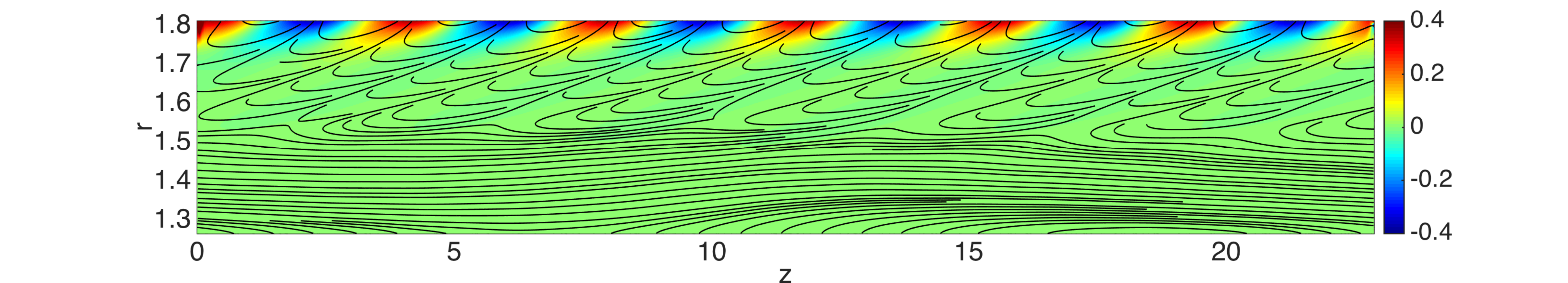} }
\centerline{\includegraphics[width=16 cm]{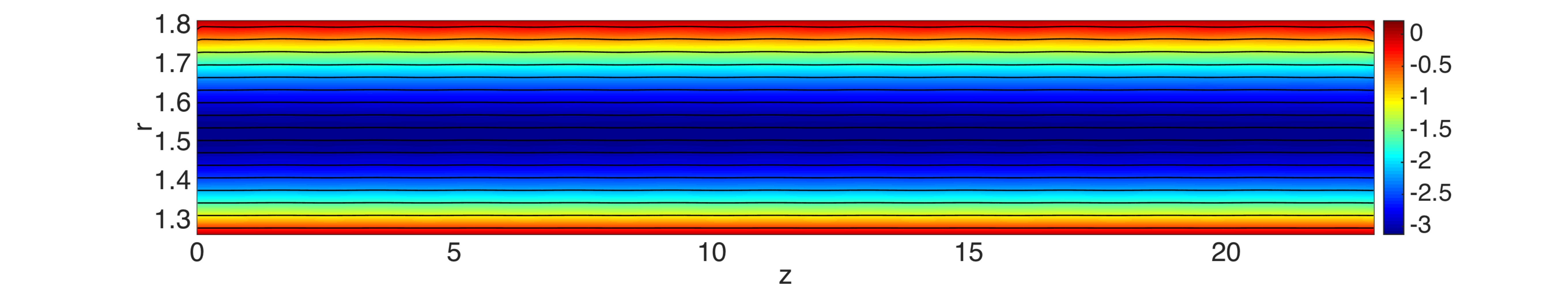} }

\caption{Structure of the axial velocity (top) and the radial magnetic
  field (bottom) for $Rm=120$, $Ha=100$ and $Re=100$, after the
  transition ($P_{ext}=-0.6$). Both velocity streamlines and magnetic
  field lines are indicated.}
\label{Poiseuille}
\end{figure*}

When the maximum of $F_B$ is reached, the system falls on the
ascending part of the curve and becomes unstable: on this branch, any
perturbation in which the bulk flow slows down implies an increase of
the difference of speed between the travelling field and the
fluid, and therefore leads to a stronger shear of the magnetic field
lines. As a consequence, the magnetic field is expelled away from the
bulk flow and the total {\it accelerating} Lorentz force decreases,
which in turns enhanced the initial braking of the fluid. The magnetic
tension cannot counteract the pressure gradient anymore, which amplifies the initial perturbation: the system rapidly moves along the
ascending branch of its characteristic until another source of
dissipation comes into play. This is the saddle-node bifurcation predicted by the model described in section 2. In our simulations, the instability stops
when the viscosity of the fluid counteracts the effect of the adverse
pressure gradient, thus leading to the Poiseuille-type regime (corresponding to the solution $A$ in Fig. \ref{PQtheory}).

Figure \ref{Poiseuille} shows colorplots of both the radial magnetic
field (top) and the axial velocity (bottom) after the saturation of
this instability. The top figure clearly illustrates the magnetic
field expulsion. First, compared to Fig. \ref{lamstruc}, the magnetic
field lines are clearly tilted by the axial velocity and reconnect
into a region very close to the outer cylinder, where the electrical
currents are imposed. Second, the colorplot also shows that most of
the magnetic energy is expelled from the bulk of the flow and confined
close to the outer boundary. The corresponding skin depth is then controlled by the slip between the velocity of the fluid and the  wave speed. In this region, the induced magnetic
field is then strongly tilted in the streamwise direction, which
explains why previous Hartmann-type boundary layer are not generated
anymore at the boundaries. Fig. \ref{Poiseuille}-bottom shows the corresponding Poiseuille flow.

Note that the above scenario also applies
to configurations in which no pressure gradient is applied, like the
simulations described in the previous section. In these simulations,
the adverse pressure gradient simply originates from the pressure loss
due to viscous friction, and the instability occurs by increasing $Rm$
at fixed pressure gradient (while it is the opposite in the present
section).\\

It is now interesting to relate this {\it flux expulsion} scenario to the predictions that can be made for the occurence of the instability. Since flux expulsion is in fact related to the difference of speed between the fluid and the TMF, it is more convenient to describe the behavior of the flow in the reference frame of the TMF, where the velocity of the fluid is $V=(u-c)$. Stalling of the flow occurs when the magnetic tension can no longer stabilize the system, i.e. when the typical spatial variations of the magnetic field {\it in the bulk flow} are strong enough, $\delta B/B_0>1$. Two different choices can be made on the typical scale of $\delta B$, depending on the force balance in the Navier-Stokes equation. If the flow is laminar, the magnetic tension $\frac{1}{\mu_0}(B.\nabla) B$ can be balanced with the viscous force, such that $\delta B\sim \mu_0\rho\nu V/B_0l$, leading to the following criterion for the occurence of flux expulsion:

\begin{equation}
M_b>\sqrt{Re_s} 
\label{MbRe}
\end{equation}

\noindent where $Re_s=(u-c)l_0/\nu=Re(1-q)$ is the kinetic Reynolds number based on the slip. The dimensionless number $M_b=(u-c)\sqrt{\mu_0\rho}/B_0$ is the ratio
between the fluid velocity (in the frame of reference of the travelling magnetic field)  and the celerity of Alfven waves. As a consequence, $M_b$ can be regarded as a {\it magnetic}  or {\it Alfvenic Mach number}, which is an
indication of the tendency of the magnetic field to be expelled by the flow (against the restoring magnetic tension).

Note that when applied to large scale electromagnetic pumps, the criterion \ref{MbRe} appears to be extremely restrictive: for a turbulent flow ($Re\sim 10^6$) of liquid sodium flowing at $10$m.s$^{-1}$, instability may be expected only if $B_0<1$Gauss. However, for turbulent flows, it is more reasonable to balance the magnetic force with inertia, giving  $\delta B\sim \mu_0\rho V^2/B_0$. In this case, the condition for instability simply reduces to:

\begin{equation}
M_b> 1
\label{Mb1}
\end{equation}

This new condition, only valid for turbulent flows, is different from the classical prediction $Rm_s>1$ of the simple block velocity model presented in the first section of this article, and predicts flux expulsion for much larger magnetic field than the viscous scaling discussed above. It provides an interesting new criterion, very simple, for predicting the
occurrence of the stalling instability. An important point is that this criterion is local, independently of the size of the system. 

\begin{figure}
\centerline{\includegraphics[width= 8 cm]{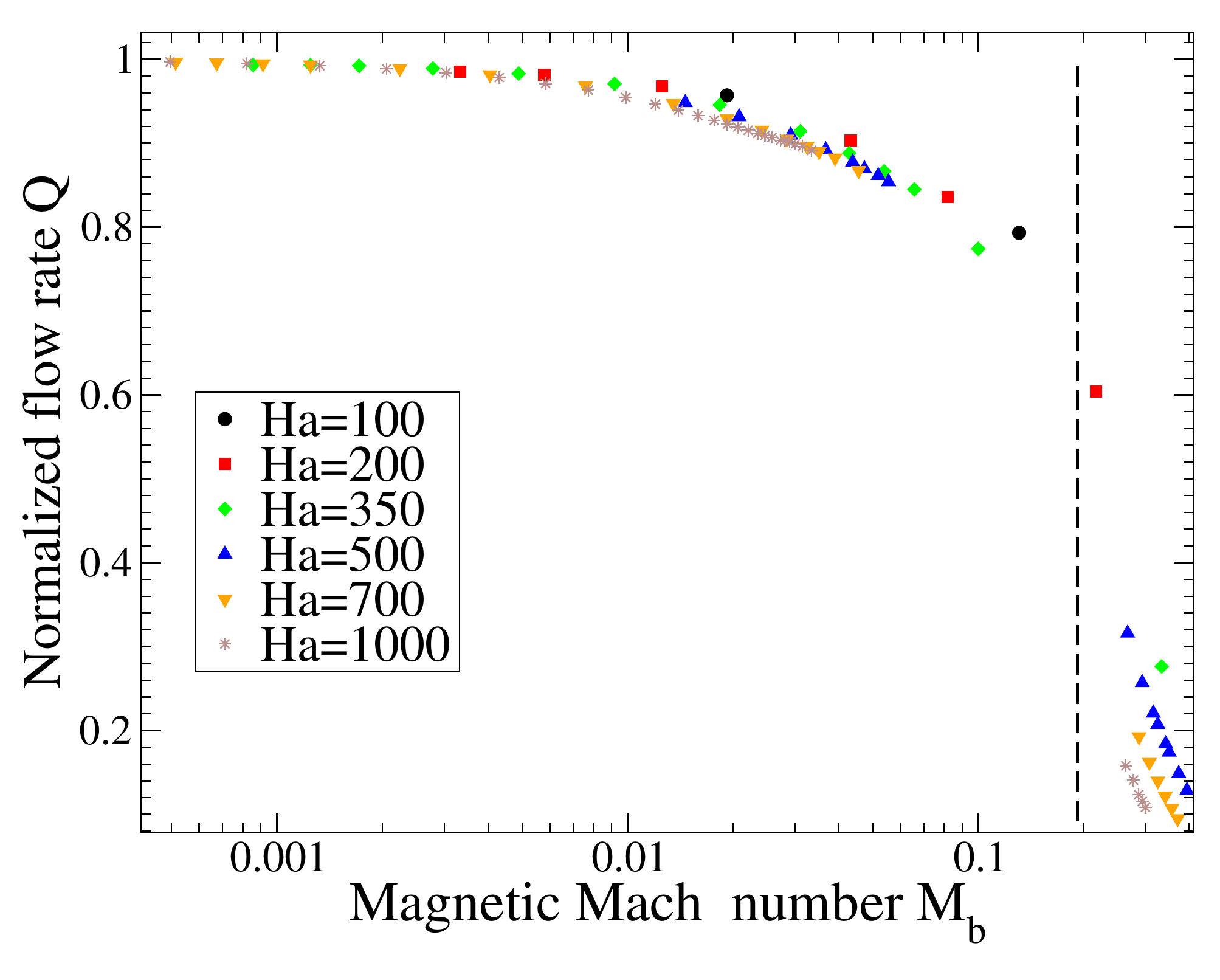} }              
\caption{Bifurcation of the axial flow rate as a function of the magnetic Mach number $M_{b}$, for
  different values of $Ha$. Note the rescaling of the points and the sharp transition observed at a given value of $M_{b}$.}
\label{bifM}
\end{figure}

 Fig. \ref{bifM} shows the evolution of the normalized flow rate as a function of $M_{b}$. Compared to Fig. \ref{bifRm}, the use of this new dimensionless number provides a very good rescaling of our data, suggesting that the Magnetic Mach number is more prone to predict the occurence of the stalling instability. The collapse of the points naturally disappears after the transition, where a non-magnetized Poiseuille regime is instead achieved.
Despite the small values of  $Re/Ha$ in our simulations, the transition seems to follows prediction of eq. (\ref{Mb1}) rather than eq. (\ref{MbRe}). This suggests that the laminar/turbulent transition occurs for a kinetic Reynolds number much smaller than the one generally observed in Hartmann flows, and follows a different scaling.

Finally, note that the above discussion using the Magnetic Mach number concerns the transition from synchronous to stalled flows, but implicitly suppose a large magnetic Reynolds number. Indeed, the hysteresis related to the saddle-node bifurcation is only observed for large magnetic Reynolds number: in order to observe a discontinuous jump of the flow rate at $M_b \sim 1$, it is therefore necessary to have also $Rm_s>1$ at the transition. 

\section{Discussion and conclusion}
\label{sec:conclu}

We reported a numerical study of the MHD flow driven
by a travelling magnetic field (TMF) in an annular channel. This configuration
aims to reproduce the flow of liquid metals generated inside annular
linear induction electromagnetic pumps. The purpose of this first part
was to compare laminar simulations to some simple theoretical
predictions, and to clarify the physical mechanisms by which such
flows could become unstable at large magnetic Reynolds numbers. We have
shown that at sufficiently large $Rm$ and $Ha$, the flow undergoes a
transition similar to the stalling of an asynchronous motor. For the
small values of the kinetic Reynolds number used here, we have shown
that several aspects of the theoretical description based on an homogeneous velocity
remains valid: this model is able to correctly predict the shape of
the PQ characteristic, and the amplitude of both velocity and magnetic
field.\\

We have also identified magnetic flux expulsion as the physical
mechanism by which the instability occurs. This is particularly clear
when an external pressure gradient is applied to the system, and a
sharp transition from Hartmann-like synchronous flows to
Poiseuille-type counter-flows is observed.\\

It is interesting to note that the behavior observed in these
simulations presents strong analogies with two analytical models
proposed by \cite{Gimblett79} and by \cite{Moffatt82}.  In the first
one, the rotation of an electrically conducting solid disk in a
transverse magnetic field is investigated. The existence of a
forbidden band of rotation rates for a given driving torque is
predicted.  In the second article, the authors study a
pressure-driven flow along a rectangular channel in the presence of a
(steady) applied magnetic field which is periodic in the streamwise
direction. They find that for a critical pressure gradient, a runaway
situation is observed in which the flow rapidly increases to larger
values. This prediction has been recently confirmed by direct numerical simulations performed in a plane channel flow \cite{Bandaru15}.  Our calculations and the interpretation given in the previous
section suggest that the exact same mechanism occurs in the case of a
linear induction electromagnetic pump: as the difference of speed
between the fluid and the wave increases, magnetic field is expelled
outside the conductor and leads to a stalled regime dominated by
viscosity.

Moreover, when the aforementioned studies are transposed to our
problem, these calculations imply that if an {\it accelerating}
pressure gradient is applied to an induction pump, a transition {\it
  opposite} to the stalling instability should be observed for
sufficiently large pressure gradient: the flow bifurcates from an
Hartmann-type regime close to synchronism, toward a much faster
Poiseuille flow ultimately controlled by viscous dissipation. This
scenario has indeed been observed in our simulations (the transition
reported in Fig. \ref{decrochA} for $P_{ext}>1.25$).

 In both cases (stalling or acceleration), magnetic flux expulsion
 associated with reconnection of field lines is responsible for the
 occurrence of the instability. This interpretation allows us to
 suggest that the Magnetic Mach number $M_b$ defined in the last
 section is controlling the generation of these instabilities.\\

Finally, it is crucial to note that industrial pumps and experimental
facilities are much more complex than the simple simulations reported
here. Among the most important differences, one is related to the
presence of inlet/outlet boundary conditions, and the other to the
very large fluid Reynolds numbers involved, which implies highly turbulent
flows. In a second article \cite{Rodriguez16}, we will therefore report numerical
simulations done at larger Reynolds numbers and with more realistic
axial boundary conditions. We will show  how the bifurcation described here leads to inhomogeneous flows at larger fluid Reynolds number. 
\\

\begin{acknowledgments}
This work was supported by funding from the French program "Retour Postdoc" managed by Agence Nationale de la Recherche (Grant ANR-398031/1B1INP), and the DTN/STPA/LCIT of Cea Cadarache.
The present work benefited from the computational support of the HPC resources of GENCI-TGCC-CURIE (Project  No.   t20162a7164)  and  MesoPSL  financed  by  the  Region
Ile  de  France  where  the present numerical simulations have been performed
\end{acknowledgments}


\end{document}